\documentclass[12pt,english,english,reqno,amsfonts]{article}
\usepackage[T1]{fontenc}
\usepackage[latin1]{inputenc}
\usepackage{a4wide}
\usepackage{color}
\usepackage{graphicx}
\usepackage{amssymb}

\makeatletter

\providecommand{\LyX}{L\kern-.1667em\lower.25em\hbox{Y}\kern-.125emX\@}
\newcommand{\boldsymbol}[1]{\mbox{\boldmath $#1$}}

\usepackage{amsmath,amsfonts,amssymb,graphicx}
\numberwithin{equation}{section}

\def\Re{\mbox{Re}}
\def\Im{\mbox{Im}}

\usepackage{babel}
\makeatother
\begin{document}
\begin{center}{\LARGE Excited Boundary TBA}\\
 {\LARGE in the Tricritical Ising Model}%
\footnote{Talk given by F. Ravanini at the 6th Landau Workshop {}``CFT and
Integrable Models'' - Chernogolovka (Moscow), Russia, 15-21 September
2002}
\end{center}{\LARGE \par}

\bigskip{}
\begin{center}{\large G. Feverati$^{\star }$, P. A. Pearce$^{\star }$
and F. Ravanini$^{\diamond }$}\end{center}{\large \par}
\bigskip{}

\begin{center}$^{\star }$\emph{Dept. of Mathematics and Statistics,
University of Melbourne}, \emph{Australia}\\
$^{\diamond }$\emph{I.N.F.N. - Sezione di Bologna, Italy}\end{center}

\begin{abstract}
\noindent By considering the continuum scaling limit of the $A_{4}$
RSOS lattice model of Andrews-Baxter-Forrester with integrable boundaries,
we derive excited state TBA equations describing the boundary flows
of the tricritical Ising model. Fixing the bulk weights to their critical
values, the integrable boundary weights admit a parameter $\xi $
which plays the role of the perturbing boundary field
$\varphi _{1,3}$ and induces the renormalization
group flow between boundary fixed points. The boundary TBA equations
determining the RG flows are derived in the $\mathcal{B}_{(1,2)}\to \mathcal{B}_{(2,1)}$
example. The induced map between distinct Virasoro characters of the
theory are specified in terms of distribution of zeros of the double
row transfer matrix.
\end{abstract}

\section{Introduction}

Quantum Field Theories with a nontrivial boundary have received recently
a lot of attention, due to their applications in Condensed Matter,
Solid State Physics and String Theory (D-branes). A problem of great
interest is the Renormalization Group flow between different boundary
fixed points of a CFT that remains conformal in the bulk. Many interesting
results have been achieved, and flows have been studied for minimal
models of Virasoro algebra and for $c=1$ CFT (see e.g. \cite{TIM}
and references therein). Numerical scaling functions for the flow
of states interpolating two different boundary conditions can be systematically
explored by use of the Truncated Conformal Space Approach \cite{TCSA,DPTW}.

A beta function can be defined for the boundary deformations, much
the same as for the bulk perturbations of conformal field theories
\cite{gthm}. The conformal boundary conditions can thus play the
role of ultraviolet (UV) and infrared (IR) points of the flow. One
gets out of the UV point by perturbing with a relevant boundary operator
and gets into an IR one attracted by irrelevant boundary operators.

Among the possible boundary perturbations of a CFT there are some
that keep an infinite number of conservation laws. These are referred
to as \emph{integrable} boundary perturbations. In this particular
case experience suggest that there should be some exact method of
investigation, based on transfer matrix and Bethe ansatz techniques.
One of the most celebrated of these methods is the Thermodynamic Bethe
Ansatz (TBA) \cite{YangYang,Zam90} giving a set of non-linear coupled
integral equations governing the scaling functions along the RG flow.
In the following we illustrate how one can obtain TBA equations from
a lattice construction for the integrable boundary flows.

The simplest nontrivial example to consider is the second model in
the Virasoro minimal series, the Tricritical Ising Model (TIM) with
central charge $c=\frac{7}{10}$. It has interesting applications
in Solid State Physics and Statistical Physics. Its Kac table is given
here below:

\vspace{0.3cm}
\begin{center}\includegraphics{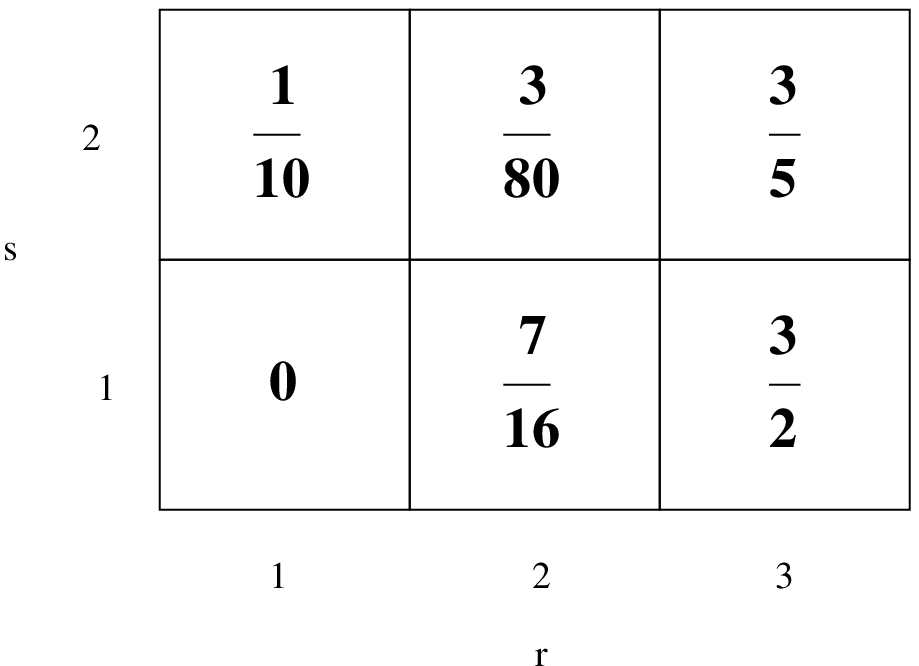}\end{center}
\vspace{0.3cm}

We have drawn only half of the table, by taking into account the well
known $\mathbb{Z}_{2}$ symmetry $(r,s)\equiv (4-r,5-s)$ that reduces
the independent chiral primaries to 6 in the TIM.

Conformal boundary conditions for minimal models with diagonal modular
invariant partition function of type $(A_{p},A_{p'})$ are classified
to be in one to one correspondence with chiral primary fields in the
Kac table. So there are 6 different conformal boundary conditions
$B_{(r,s)}$, $r=1,2,3$ and $s=1,2$ for the TIM $(A_{3},A_{4})$.

Let us consider in the two-dimensional plane $(x,t)$ the TIM defined
on a strip of thickness $L$, i.e in the region $0\leq x\leq L$ and
$-\infty <t<+\infty $, with non-trivial b.c. $B_{(r,s)}$ on the
left ($x=0$) and $B_{(r',s')}$ on the right ($x=L$) edge respectively.
This situation will be denoted $B_{(r,s)|(r',s,)}$. In the following
we are interested in boundaries of type $(r,s)|(1,1)$. We shall denote
for short $B_{(r,s)|(1,1)}=\mathcal{B}_{(r,s)}$. For this particular
class of boundary conditions the partition function simply reduces
to one single character \cite{Cardy} \[
Z_{\mathcal{B}_{(r,s)}}(q)=\chi _{(r.s)}(q)\]
where $\chi _{(r,s)}(q)$ denotes the character of the irreducible
Virasoro representation labeled by $(r,s)$ at central charge $c=\frac{7}{10}$.

For each given boundary condition $\mathcal{B}_{(r,s)}$ there is
a set of boundary operators $\phi _{(u,v)}$ that live on the edge.
If we want to keep the b.c. unchanged all along the edge, these operators
must be restricted to the conformal families appearing in the OPE
fusion of the Virasoro family $(r,s)$ with itself: $(u,v)\in (r,s)\times (r,s)$.
They are distinguished in terms of their conformal dimensions $h_{r,s}$
as relevant ($h_{r,s}<1$) and irrelevant ($h_{r,s}>1$). Of course
only relevant perturbations break scale invariance at the boundary
in such a way to get out of the fixed boundary point of a specific
boundary condition and flow to another one. So if we want to consider
possible relevant boundary perturbations of the TIM, i.e. QFT's described
by the action\[
S=S_{(r,s)}+\lambda \int _{-\infty }^{+\infty }dt\, \phi _{(u,v)}(x=0,t)\]
 -- where $S_{(r,s)}$ denotes the action of TIM with boundary condition
$\mathcal{B}_{(r,s)}$ -- we have to restrict to the following possibilities

\begin{center}\begin{tabular}{|c|c|c|}
\hline 
\multicolumn{2}{|c|}{boundary condition}&
boundary perturbations\\
\hline
\hline 
$\mathcal{B}_{(1,1)}\equiv \mathcal{B}_{(3,4)}$&
$-$&
none\\
\hline 
$\mathcal{B}_{(2,1)}\equiv \mathcal{B}_{(2,4)}$&
$0$&
none\\
\hline 
$\mathcal{B}_{(3,1)}\equiv \mathcal{B}_{(1,4)}$&
$+$&
none\\
\hline 
$\mathcal{B}_{(1,2)}\equiv \mathcal{B}_{(3,3)}$&
$-0$&
$\phi _{(1,3)}$\\
\hline 
$\mathcal{B}_{(2,2)}\equiv \mathcal{B}_{(2,3)}$&
$d$&
$\phi _{(1,2)},\, \phi _{(1,3)}$\\
\hline 
$\mathcal{B}_{(3,2)}\equiv \mathcal{B}_{(1,3)}$&
$0+$&
$\phi _{(1,3)}$\\
\hline
\end{tabular}\end{center}

Actually there are two physically different flows for each {}``pure''
perturbation (i.e. containing only one operator $\phi _{(12)}$ or
$\phi _{(13)}$), flowing to two possibly different IR destinies.
This is achieved by taking different signs in the coupling constant
in the case of $\phi _{(13)}$ flows, and real or purely imaginary
coupling constant in case of $\phi _{(12)}$ ones. The boundary condition
$\mathcal{B}_{(2,2)}$ can be perturbed by any linear combination
of the fields $\phi _{(1,2)}$ and $\phi _{(1,3)}$. The symbols $+,-,0+,-0,0,d$
represent other ways to denote the TIM conformal boundary conditions
present in the literature \cite{chim,Affleck}, and we report them
here only to facilitate translations to the reader.

Along the flow, going from UV to IR, the boundary entropy associated
to each $\mathcal{B}_{(r,s)}$\[
g_{(r,s)}=\left(\frac{2}{5}\right)^{1/4}\frac{\sin \frac{\pi r}{4}\sin \frac{\pi s}{5}}{\sqrt{\sin \frac{\pi }{4}\sin \frac{\pi }{5}}}\]
decreases \cite{gthm}. Therefore we expect to have flows only between
boundary conditions where the starting conformal boundary entropy
is higher than the final one. 

\begin{figure}
\includegraphics{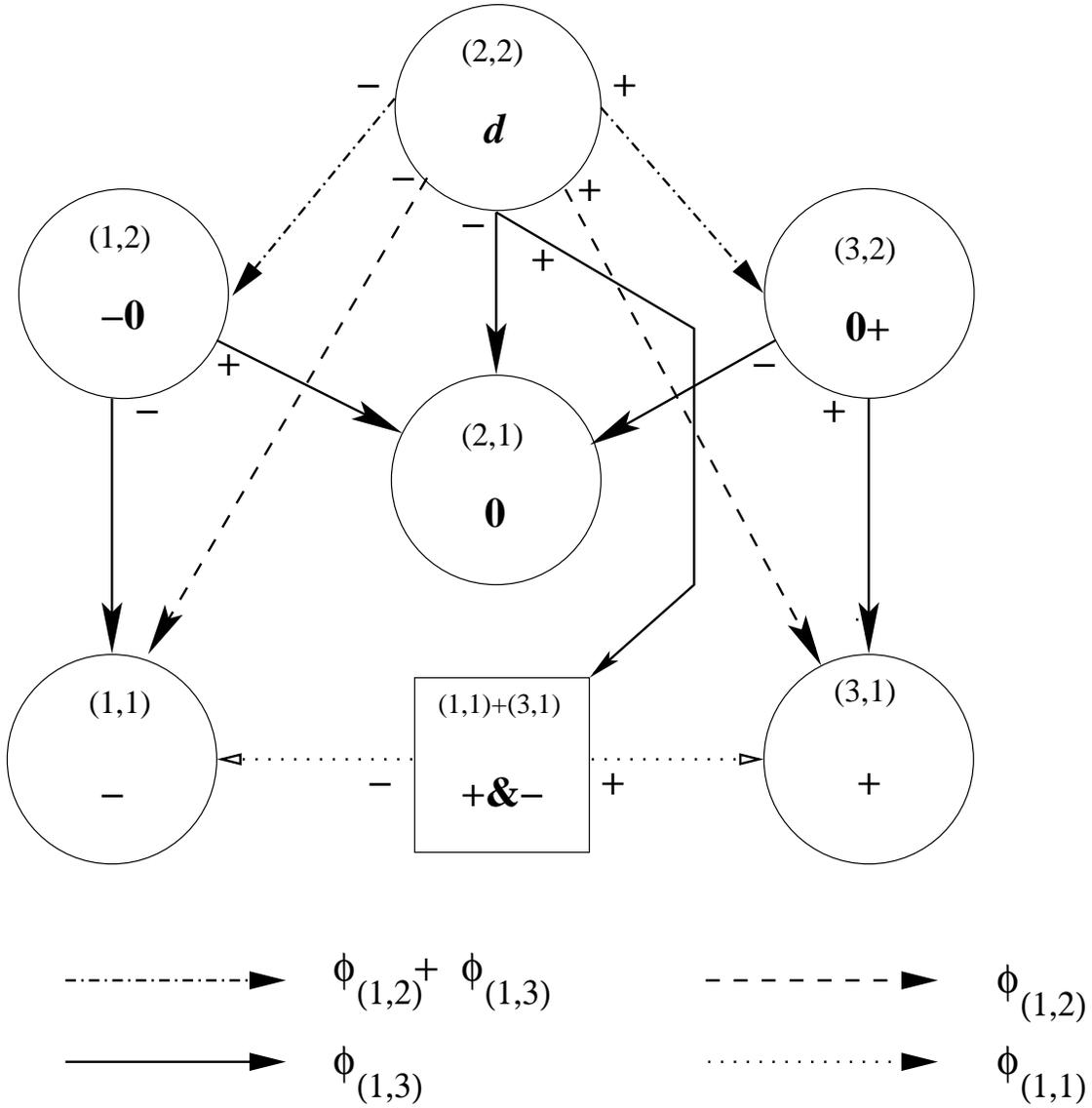}

\caption{\label{fig-flows}The boundary flows between TIM conformal b.c. Pure
$\phi _{(1,3)}$ and $\phi _{(1,2)}$ flows are integrable. More information
can be found in \cite{Affleck}.}
\end{figure}

The possible conformal boundary conditions have been studied extensively
by Chim \cite{chim} and the flows connecting them by Affleck \cite{Affleck}.
The picture can be summarized as in fig. \ref{fig-flows}. Integrability
can be investigated in a manner similar to the bulk perturbations
and it turns out that the flows generated by pure $\phi _{(1,3)}$
and $\phi _{(1,2)}$ perturbations are integrable. Instead, the flow
starting at $B_{(2,2)}$ as a perturbation which is a linear combination
of $\phi _{(1,2)}$ and $\phi _{(1,3)}$ is strongly suspected to
be non-integrable.

Notice the $\mathbb{Z}_{2}$ symmetry of fig. \ref{fig-flows}, which
is strictly related to the supersymmetry of the TIM. Investigation
of the supersymmetric aspects of the TIM boundary flows is, however,
out of the scope of the present talk.

We deduce TBA equations for the integrable flows from the functional
relations for the transfer matrix of a lattice RSOS realization of
the model. We will focus, as an example, on the flow obtained by perturbing
the $B_{(1,2)}$ boundary TIM by the relevant boundary operator $\varphi _{(1,3)}$\begin{equation}
UV=\mathcal{B}_{(1,2)}\mapsto \mathcal{B}_{(2,1)}=IR,\qquad \chi _{1,2}(q)\mapsto \chi _{2,1}(q)\label{protoflow}\end{equation}
 This flow induces a map between Virasoro characters $\chi _{r,s}(q)$
of the theory, where $q$ is the modular parameter. The physical direction
of the flow from the ultraviolet (UV) to the infrared (IR) is given
by the relevant perturbations and is consistent with the $g$-theorem~\cite{gthm}.
The approach outlined here, however, is quite general and should apply,
for example, to all integrable boundary flows of minimal models.

\section{Scaling of Critical $A_{4}$ Model with Boundary Fields}

It is well known that the TIM is obtained as the continuum scaling
limit of the generalized hard square model of Baxter~\cite{HardSq}
on the Regime~III/IV critical line, \emph{i.e.} the $A_{4}$ RSOS
integrable lattice model of Andrews-Baxter-Forrester~\cite{ABF},
with critical Boltzmann weights

\begin{center}\textcolor{red}{\includegraphics{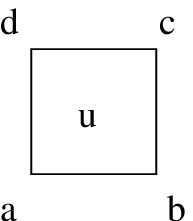}}\textcolor{black}{\[
=W\left.\left(\begin{array}{cc}
 d & c\\
 a & b\end{array}
\right|u\right)=\frac{\sin (\frac{\pi }{5}-u)}{\sin \frac{\pi }{5}}\delta _{a,c}+\frac{\sin u}{\sin \frac{\pi }{5}}\left(\frac{\sin \frac{\pi a}{5}\sin \frac{\pi c}{5}}{\sin \frac{\pi b}{5}\sin \frac{\pi d}{5}}\right)^{1/2}\delta _{b,d}\]
}\end{center}

\noindent satisfying the Yang-Baxter equation. The indices $a,b,c,d=1,...4$
take values on the $A_{4}$ Dynkin diagrams, whose adjacency dictates
which are the nonzero elements of the Boltzmann weights.

In the presence of integrable boundaries, one must also introduce
the so called boundary Boltzmann weights, satisfying the Boundary
Yang-Baxter equations. Their general form has been given in \cite{BP},
for our purposes here we select 

\begin{center}\textsf{\textcolor{black}{\includegraphics{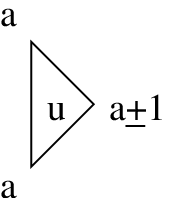}\[
=K\left(\left.\begin{array}{c}
 a\\
 a\end{array}
\, a\pm 1\right|u\right)=\left(\frac{\sin \frac{\pi (a\pm 1)}{5}}{\sin \frac{\pi a}{5}}\right)^{1/2}\frac{\sin (u\pm \xi _{L})\sin (u\mp \frac{\pi a}{5}\mp \xi _{L})}{\sin ^{2}\frac{\pi }{5}}\]
}}\end{center}

\noindent and

\begin{center}\textsf{\textcolor{black}{\includegraphics{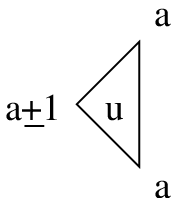}\[
=K\left(\left.a\pm 1\, \begin{array}{c}
 a\\
 a\end{array}
\right|u\right)=\left(\frac{\sin \frac{\pi (a\pm 1)}{5}}{\sin \frac{\pi a}{5}}\right)^{1/2}\]
}}\end{center}

\noindent All these Boltzmann weights are periodic under $u\to u+\pi $.
The scaling energies of the TIM are obtained~\cite{OPW} from the
scaling limit of the eigenvalues of the commuting double-row transfer
matrices $\boldsymbol {D}(u)$~\cite{BPO}

\begin{center}\includegraphics{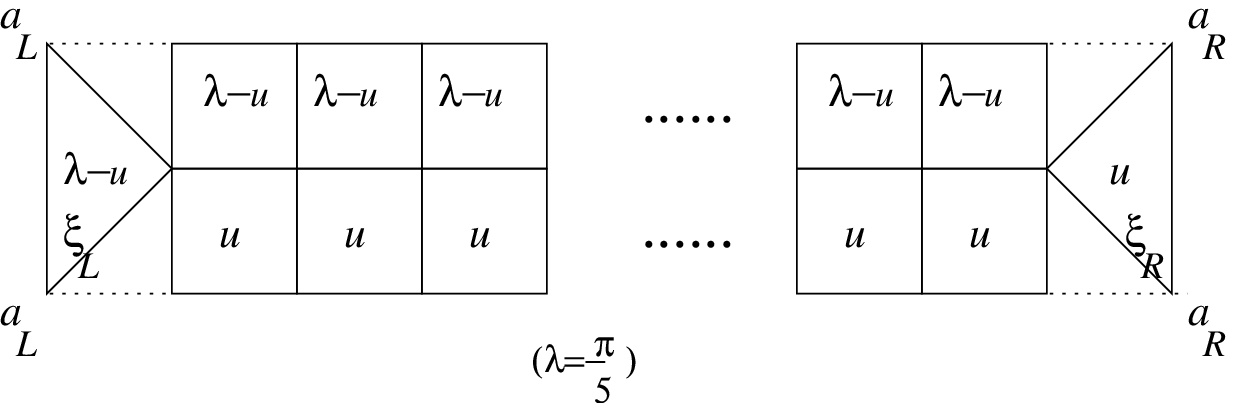}\end{center}

\noindent of the $A_{4}$ lattice model with $N$ faces in a row.
Integrability means $[\mathbf{D}(u),\mathbf{D}(v)]=0$, $\forall u,v\in \mathbb{C}$.
The physical region is given by $0<u<\frac{\pi }{5};$ crossing symmetry
$\mathbf{D}(u)=\mathbf{D}(\frac{\pi }{5}-u)$ implies that the isotropic
point is $u=\frac{\pi }{10}$. The matrix $\mathbf{D}(u)$ satisfies
the property\[
\dim \, \mathbf{D}_{(a_{L},a_{R})}^{(N)}=(\mathcal{A}^{N})_{a_{L},a_{R}}\]
where the matrix $\mathcal{A}$ is the incidence matrix of the $A_{4}$
Dynkin diagram. Here the notation $\mathbf{D}_{(a_{L},a_{R})}$ means
the sub-matrix of $\mathbf{D}$ for fixed $a_{L},a_{R}$ boundary
spins. It is convenient to introduce the \emph{normalized} double
row transfer matrix \[
\mathbf{t}(u)=\mathcal{P}(u,\xi _{L})\mathbf{D}(u)\]
 that satisfies the following functional equation:\begin{equation}
\mathbf{t}(u)\mathbf{t}\left(u+\frac{\pi }{5}\right)=1+\mathbf{t}\left(u+\frac{3\pi }{5}\right).\label{func-equat}\end{equation}

\noindent For the interested reader, the expression for the normalization
factor $\mathcal{P}(u,\xi _{L})$ is given in \cite{OPW}. Even if
the equation for $\mathbf{t}$ looks boundary independent, the solutions
are fixed by the behavior of the zeros, that are boundary dependent.
The transfer matrix $\mathbf{D}(u)$ is entire and has only zeros;
$\mathbf{t}(u)$ has instead two complex conjugate poles (coming from
$\mathcal{P}(u,\xi _{L})$) in the strip $-\frac{\pi }{10}<\Im u<\frac{3\pi }{10}$.
We shall see that they will affect the boundary driving term in the
TBA equations we are going to deduce.

In a lattice of $N$ horizontal and $M$ vertical sites, with periodic
boundary conditions on the vertical direction and $a_{L},a_{R}$ boundary
conditions on the horizontal one, the partition function of the tricritical
hard square model can be written as\[
Z_{MN}(u)=\mathrm{Tr}_{\mathcal{H}}\left(\mathbf{D}_{(a_{L},a_{R})}(u)^{M}\right)\sim e^{-NMf_{bulk}(u)-Mf_{surf}(u)}Z(q)\]
where $f_{bulk}$ and $f_{surf}$ are the bulk and surface contributions
to the free energy respectively. $Z(q)$ is the universal conformal
partition function, with modular parameter\[
q=e^{-2\pi \frac{M}{N}\sin 5u}\]
Therefore the behavior of each transfer matrix eigenvalue is \[
\log d(u)=-Nf_{bulk}(u)-f_{surf}(u)-\frac{2\pi }{N}\left(\Delta -\frac{c}{24}+n\right)\sin 5u+o\left(\frac{1}{N}\right)\]
with $n$ a nonnegative integer. It is then logical to consider for
the eigenvalues $t(u)$ of $\mathbf{t}(u)$, the decomposition for
$N\to \infty $\[
t(u)=f(u)g(u)l(u)\]
where $f(u)\sim O(N)$, $g(u)\sim O(1)$ and $l(u)\sim O(1/N)$.

In general the conformal partition function expands over characters
of the $c=\frac{7}{10}$ Virasoro algebra\[
Z(q)=\sum _{r,s}\mathcal{N}_{r,s}\chi _{r,s}(q)\]
where $\mathcal{N}_{r,s}$ are nonnegative integers describing the
multiplicities of Virasoro representations in the Hilbert space.

\begin{figure}
\begin{center}\includegraphics{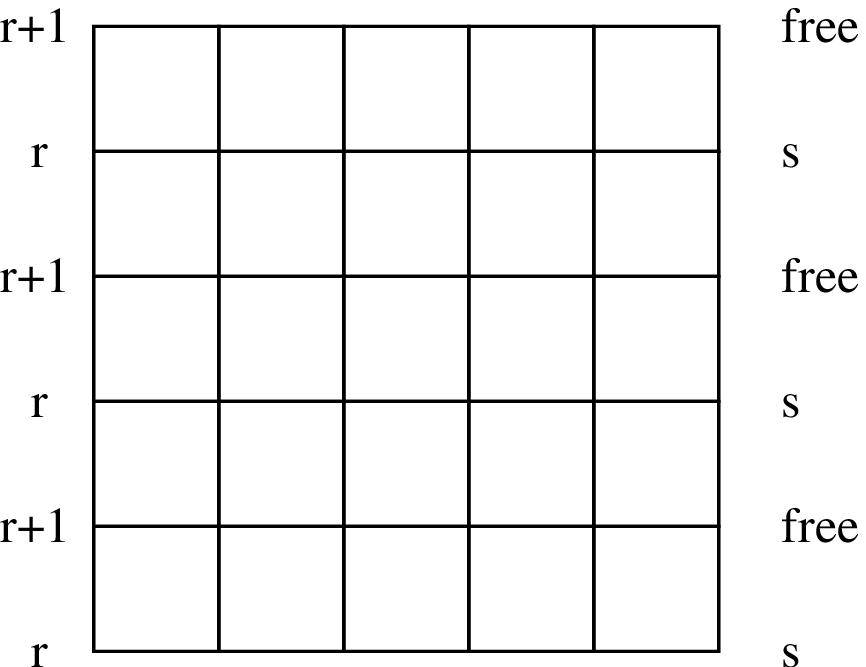}\end{center}

\begin{center}Figure 2: \label{cap:lattice}Lattice boundary configuration
leading to Cardy partition function\end{center}
\end{figure}

Bauer and Saleur \cite{bauer-saleur} used Coulomb gas techniques
to show that if one chooses the boundary conditions leading to the
configuration depicted in fig.\ref{cap:lattice} the single character
$\chi _{(r,s)}$, i.e. the partition function of TIM with $\mathcal{B}_{(r,s)}$
boundary conditions, can be realized in the scaling limit $N\to \infty $.
The configuration can be built by taking the boundary Boltzmann weights
described above, with $a_{L}=r$, $a_{R}=s$ and $\Re \xi _{L}=\frac{\pi }{10}$.
This latter choice kills the weight alternating $r$ and $r-1$ on
the left at the isotropic point $u=\frac{\pi }{10}$.

At criticality there exist integrable boundary conditions associated
with each conformal boundary condition $(r,s)$ carrying 0,1 or 2
arbitrary boundary parameters $\xi _{L},\xi _{R}$~\cite{BP}. 
To consider boundary flows we must fix the bulk at its critical point
and vary the boundary parameters $\xi _{L},\xi_R$. However, these
{}``fields'' are irrelevant in the sense that
they do not change the scaling energies if they are real and restricted
to appropriate intervals~\cite{OPW}. Explicitly, the cylinder partition
functions obtained~\cite{OPW,BP} from $(r,s)$ integrable boundaries
with $\xi _{L},\xi_{R}$ real are independent of $\xi _{L},\xi _{R}$
and given by single Virasoro characters \begin{equation}
\mathcal{B}_{(r,s)}:\qquad Z(q)=\chi _{r,s}(q).\end{equation}
 The reason for this is that in the lattice model the fields $\xi _{L}$
and $\xi_{R}$ control the location of zeros of eigenvalues of $\mathbf{D}(u)$
on the real axis in the complex plane of the spectral parameter $u$
whereas only the zeros in the scaling regime, a distance $i\log N$
out from the real axis, contribute in the scaling limit $N\rightarrow \infty $.
The solution is to scale the imaginary part of $\xi _{L},\xi_{R}$
as $\log N$. This is allowed because $\xi _{L},\xi_{R}$ are arbitrary
complex fields.

We find that only one parameter at the time (we call it $\xi_L$)
is needed to induce the $\varphi _{(1,3)}$ RG flows and is identified 
as the thermal boundary field:
\begin{equation}
\xi _{L}\sim \varphi _{(1,3)}.\end{equation}

For the flow (\ref{protoflow}), following \cite{OPW}, we consider
the $(r,s)=(2,1)$ boundary with the $(1,1)$ boundary on the right
(with no boundary field) and the $(2,1)$ boundary on the left with
boundary field $\xi _{L}$. We scale $\xi _{L}$ as \begin{equation}
\textrm{Re}(\xi _{L})=\frac{3}{10}\pi,\qquad \textrm{Im}(\xi _{L})=\frac{-\xi +\log N}{5}\label{scaling-xi}\end{equation}
 where $\xi $ is real. In terms of boundary weights, the $(1,2)$
boundary is reproduced in the limit $\textrm{Im}(\xi _{L})\rightarrow \pm \infty $
whereas the $(2,1)$ boundary is reproduced in the limit $\textrm{Im}(\xi _{L})\rightarrow 0$.

The scaling behavior of the $A_{4}$ lattice model at the boundary
fixed points is described by the TBA equations of O'Brien, Pearce
and Warnaar~\cite{OPW} (OPW). We generalize their analysis to include
the scaling boundary field~(\ref{scaling-xi}).

\section{Classification of zeros}

The zeros of $\mathbf{D}_{(a_{L},a_{R})}(u)$ can be classified in
two strips, the first given by $-\frac{\pi }{10}<\Re (u)<\frac{3\pi }{10}$
and the second by $\frac{2\pi }{5}<\Re (u)<\frac{4\pi }{5}.$ We refer
to them as strip I and II respectively. Each zero has to be accompanied
by its complex conjugate, and there are only two zeros on the real
axis depending on $\Re \xi _{L}$ that do not play any role in the
following. So we are led to consider only zeros in the upper half
$u$ complex plane excluding the real axis. For $N$ sufficiently
large general considerations supported by numerical inspection allow
to say that there exists

\begin{itemize}
\item $m_{1}$ 1-strings at $\frac{\pi }{10}+iy_{k}^{(1)}$ and $n_{1}$
2-strings at $-\frac{(2j+1)\pi }{10}+iz_{k}^{(1)}$with $j=0,1$ in
strip I
\item $m_{2}$ 1-strings at $\frac{3\pi }{5}+iy_{k}^{(2)}$ and $n_{2}$
2-strings at $\frac{(2j+1)\pi }{5}+iz_{k}^{(2)}$ with $j=0,1$ in
strip II
\end{itemize}
The precise classification can be done in terms of the relative ordering
(with respect to their imaginary parts) of the 1 and 2-strings. This
can be achieved by giving the sequence of numbers $I_{k}^{(i)}$ of
2-strings with $z_{k}^{(i)}>y_{k}^{(i)}$. In the case of $\mathcal{B}_{(1,2)}$
this must be supplemented \cite{OPW} by a parity of strings $\sigma =\pm 1$.
Because we order the zeros as $y_{k+1}^{(i)}>y_{k}^{(i)},$ the quantum
numbers must satisfy $I_{1}^{(i)}\geq I_{2}^{(i)}\geq ...\geq I_{m_{i}}^{(i)}\geq 0.$
Moreover, in the second strip the largest quantum number must be less
or equal the number of 2-strings of type 2, $n_{2}\geq I_{1}^{(i)}.$
The 2-strings in strip I are infinite in number at the scaling limit
(see below), and their quantum numbers are unbounded. We write this
sequence as\[
I=(I_{1}^{(1)},I_{2}^{(1)},...,I_{m_{1}}^{(1)}|I_{1}^{(2)},I_{2}^{(2)},...,I_{m_{2}}^{(2)})_{\sigma }\qquad \mathrm{with}\quad \sigma =\left\{ \begin{array}{cc}
 \pm 1 & \mathrm{for}\, \mathcal{B}_{(1,2)}\\
 0 & \mathrm{for}\, \mathcal{B}_{(2,1)}\end{array}
\right.\]
There are constraints on the numbers of strings, analyzed in detail
in \cite{OPW}, depending on the specific $\mathcal{B}_{(r,s)}$ condition

\begin{itemize}
\item for $\mathcal{B}_{(1,2)}$ ($m_{1}$ odd, $m_{2}$ even)\begin{equation}
n_{1}=\frac{N+m_{2}+\sigma }{2}-m_{1}\quad ,\quad n_{2}=\frac{m_{1}-\sigma }{2}-m_{2}\label{mn1}\end{equation}

\item for $\mathcal{B}_{(2,1)}$ ($m_{1}$ odd, $m_{2}$ odd)\begin{equation}
n_{1}=\frac{N+m_{2}}{2}-m_{1}\quad ,\quad n_{2}=\frac{m_{1}+1}{2}-m_{2}\label{mn2}\end{equation}

\end{itemize}
In the scaling limit $N\to \infty $ the number of 2-strings in strip
I becomes infinite: $n_{1}\sim O(N)$, while the number of the other
three types of roots (1-strings in both strips and 2-strings in strip
II) remains of order $O(1)$. In \cite{OPW} it was shown that the
pattern of zeros fixes the energies (here $\mathbf{m}=(m_{1},m_{2})$
and $\mathbf{C}$ is the $A_{4}$ Cartan matrix)\[
E(\mathcal{B}_{(1,2)})=\frac{1}{10}+\frac{\mathbf{mCm}}{4}+\sigma \frac{m_{1}+m_{2}}{2}+\sum _{i=1}^{2}\sum _{k=1}^{m_{i}}I_{k}^{(i)}\]
\[
E(\mathcal{B}_{(2,1)})=\frac{7}{16}-\frac{1}{2}+\frac{\mathbf{mCm}}{4}+\sum _{i=1}^{2}\sum _{k=1}^{m_{i}}I_{k}^{(i)}\]
Out of these energy expressions it is possible to reconstruct the
characters $\chi _{1,2}$ and $\chi _{2,1}$. First consider the so
called \emph{finitized characters} \cite{FinChar}\[
\chi _{r,s}^{N}(q)=q^{-\frac{c}{24}}\sum _{E}q^{E}=q^{\frac{c}{24}}{Tr}\left(\frac{\mathbf{D}^{(N)}(u)}{D_{0}^{(N)}(u)}\right)^{\frac{M}{2}}\]
where $D_{0}^{(N)}$ is the largest eigenvalue of the double row transfer
matrix with $(r,s)$ boundary conditions and $N$ sites. The limit
$N\to \infty $ of such objects can be shown to give the well known
formulae for the TIM characters (for details see \cite{FPR}).

Once the characters have been related in such a strict way to the
distribution of zeros in strips I and II, it is also possible to see
how they are mapped one another along the flow $\mathcal{B}_{(1,2)}\to \mathcal{B}_{(2,1)}$.
For this purpose it is more convenient to consider the reverse flow
$\mathcal{B}_{(2,1)}\to \mathcal{B}_{(1,2)}$ from IR to UV. The counting
of the total number of zeros in the upper $u$ plane for a $\mathcal{B}_{(2,1)}$
configuration is $N$, while for $\mathcal{B}_{(1,2)}$ is $N-1$.
Therefore one zero must disappear to infinity along this reverse flow
(or appear from infinity in the physical flow). We have found 3 mechanisms
that seem to exhaust the possible ways for such a phenomenon to be
realized \cite{FPR}

\begin{itemize}
\item A. The top 1-string in strip~2 flows to $+\infty $, decoupling from
the system while $m_{1}$, $n_{1}$ and $n_{2}$ remain unchanged.
This mechanism applies in the IR when $I_{m_{1}}^{(1)}=I_{m_{2}}^{(2)}=0$
and produces frozen states in the UV with $\sigma =1$ and $I_{m_{1}}^{(1)}=0$.
\item B. The top 2-string in strip~1 and the top 1-string in strip~2 flow
to $+\infty $ and a 2-string comes in from $+\infty $ in strip~2
becoming the top 2-string. Consequently, each $I_{j}^{(1)}$ decreases
by $1$ and each $I_{k}^{(2)}$ increases by $1$. This mechanism
applies in the IR when $I_{m_{1}}^{(1)}>0$ and $I_{m_{2}}^{(2)}=0$
and produces states in the UV with $\sigma =-1$ and either $m_{2}=0$
or $I_{m_{2}}^{(2)}>0$.
\item C. The top 2-string in strip~2 flows to $+\infty $ and a 1-string
in strip~2 comes in from $+\infty $. Consequently, each $I_{k}^{(2)}$
decreases by $1$. This mechanism applies in the IR when $I_{m_{2}}^{(2)}>0$
and produces states in the UV with $\sigma =-1$ and $I_{m_{2}}^{(2)}=0$. 
\end{itemize}
Observe that 

\begin{itemize}
\item these mappings are in fact one-to-one
\item the 1-strings in strip~1 are never involved
\item the change in parity of $m_{2}$ is consistent with (\ref{mn1}) and
(\ref{mn2}).
\item for A,B: $m_{2}^{UV}=m_{2}^{IR}-1$ and for C: $m_{2}^{UV}=m_{2}^{IR}+1$.
\end{itemize}
By doing such substitutions $m_{2}^{UV}\to m_{2}^{IR}\pm 1$ in the
finitized characters, it is possible to show explicitly that $\chi _{1,2}(q)\to \chi _{2,1}(q)$
under the physical flow $\mathcal{B}_{(1,2)}\to \mathcal{B}_{(2,1)}$.
See \cite{FPR} for details.

\section{Derivation of Boundary TBA Equations}

\subsection{On the lattice}

The recurrence relation (\ref{func-equat}) for the $O(N)$ piece
yields the following functional equation for the function $f(u)$\[
f(u)f(u+\frac{\pi }{5})=1\]
Plugging its solution (for details see \cite{OPW}) back into the
functional equation (\ref{func-equat}) it yields an equation for
$g(u)$ that in turn can be solved in the first analyticity strip
$-\frac{\pi }{10}<\Re u<\frac{\pi }{10}$. In the following, for any
$h(u)=f(u),g(u),l(u)$ we introduce the functions\[
h_{1}(x)=h\left(\frac{\pi }{10}+i\frac{x}{5}\right)\quad ,\quad h_{2}(x)=h\left(\frac{3\pi }{5}+i\frac{x}{5}\right)\]
each one being analytic in the corresponding strip I or II. 

The recurrence relation for $g(u)$ yields \[
g_{1}(x,\xi _{L})=\tanh ^{2}\frac{x}{2}\tanh \frac{x+5\Im \xi _{L}}{2}\tanh \frac{x-5\Im \xi _{L}}{2}\]
and an analogous formula for $g_{2}(x)$ that we do not need in the
following. The lattice TBA equations contain the boundary contribution
encoded in the function $g_{1}(x)$. It becomes exactly $1$ at the
conformal points and takes $\xi _{L}$ dependent values during the
flow. Notice that $g_{1}(x)$ depends on $\xi _{L}$ only through
its imaginary part.

Inserting this value for $g_{1}(x)$ into the recurrence equation
again, one can determine a constraint to be satisfied by the last
pieces $l_{i}(x)$, in the form of an integral equation. Putting all
these things together we arrive finally to two coupled integral equations
to be satisfied by $t_{1}(x)$ and $t_{2}(x)$\begin{eqnarray}
-\log t_{1}(x) & = & -\log g_{1}(x)+\sum _{j=1}^{m_{1}}\log \tanh \frac{1}{2}(x-v_{j}^{(1)})-K*\log (1+t_{2})\label{latt-TBA1}\\
-\log t_{2}(x) & = & -N\log (\tanh ^{2}\frac{x}{2})+\sum _{k=1}^{m_{2}}\log \tanh \frac{1}{2}(x-v_{k}^{(2)})-K*\log (1+t_{1})\label{latt-TBA2}
\end{eqnarray}
 where the kernel of the convolution is:\[
K(x)=\frac{1}{2\pi \cosh (x)}.\]

Observe that in $t_{i}$ there is an implicit $N$ dependence that
will take a role performing the scaling limit. Only the 1-strings
appear explicitly in the TBA equations, $v_{k}^{(i)}$ being their
imaginary parts. Their location is dictated by the quantization contitions\begin{equation}
-\log \left(-t_{1}(v_{k}^{(2)})\right)=n_{k}^{(2)}\pi i\label{posiz1}\end{equation}
\begin{equation}
-\log \left(-t_{2}(v_{k}^{(1)})\right)=n_{k}^{(1)}\pi i\label{posiz2}\end{equation}
where $n_{k}^{(1)}=2(m_{1}-k+I_{k}^{(1)})+1-m_{2}$ and $n_{k}^{(2)}=2(m_{2}-k+I_{k}^{(2)})+1-m_{1}$.

\subsection{The scaling limit}

In the scaling limit $N\to \infty $ where $\Im \xi _{L}$ rescales
as in (\ref{scaling-xi}) we define $\hat{h}_{i}(x)=\lim _{N\to \infty }h_{i}(x+\log N)$.
We have\begin{equation}
\hat{g}_{1}(x,\xi )=\tanh \frac{x+\xi }{2}\label{ghat}\end{equation}
 with $\xi \rightarrow -\infty $ and $\xi \rightarrow +\infty $
corresponding to $\mathcal{B}_{(1,2)}$ and $\mathcal{B}_{(2,1)}$
respectively.

Eqs. (\ref{posiz1},\ref{posiz2}) fix the following large $N$ behavior
for $v^{(i)}$:\[
y=\lim _{N\, \rightarrow \, \infty }(5v-\log N)=\textrm{finite val}.\]
 Performing this scaling limit on (\ref{latt-TBA1},\ref{latt-TBA2},\ref{posiz1},\ref{posiz2}),
and introducing the so called \emph{pseudoenergies} \begin{equation}
\varepsilon _{i}(x)=-\log \left(-\hat{t}_{i}(x)\right)\label{pseudo}\end{equation}
and the functions\[
L_{i}(x)=\log (1-e^{-\varepsilon _{i}(x)})\]
we obtain the following final TBA equations: \begin{eqnarray}
\varepsilon _{1}(x) & = & -\log \tanh \frac{x+\xi }{2}-\sum _{j=1}^{m_{1}}\log \tanh \left(\frac{y_{j}^{(1)}-x}{2}\right)-K*L_{2}(x)\label{TBA-1}\\
\varepsilon _{2}(x) & = & 4e^{-x}-\sum _{k=1}^{m_{2}}\log \tanh \left(\frac{y_{k}^{(2)}-x}{2}\right)-K*L_{1}(x).\label{TBA-2}
\end{eqnarray}
 The definition of the scaling theory must be completed by the energy
formula \begin{equation}
\begin{array}{c}
 {\displaystyle E(\xi )-\frac{c}{24}=\frac{1}{\pi }\lim _{R\, \rightarrow \, 0}RE(R)=}\\
 \\ \displaystyle
 \frac{1}{\pi }\left[\sum _{j=1}^{m_{1}}2e^{-y_{j}^{(1)}}-\int _{-\infty }^{\infty }\frac{dx}{\pi }e^{-x}\log (1+e^{-\varepsilon _{2}(x)})\right]\end{array}
\label{c-tilde}\end{equation}
 where the limit $R\, \rightarrow \, 0$ means that all this computations
are performed at the critical temperature of the TIM. The boundary
parameter $\xi $ appears only implicitly in (\ref{c-tilde}). The
last piece required to complete the scaling picture is the location
of the zeros. From (\ref{posiz1},\ref{posiz2}) we obtain: \begin{eqnarray}
\varepsilon _{2}(y_{j}^{(1)}-i\frac{\pi }{2}) & = & n_{k}^{(1)}\pi i\qquad \textrm{first strip zeros}\label{location-1}\\
\varepsilon _{1}(y_{k}^{(2)}-i\frac{\pi }{2}) & = & n_{k}^{(2)}\pi i\qquad \textrm{second strip zeros}\label{location-2}
\end{eqnarray}
 The two-string zeros satisfy similar equations, using the substitution
$y_{k}^{(i)}\rightarrow z_{k}^{(i)}.$ 

An immediate consequence is that $L_{i}(x)$ vanish on the zeros of
the same strip: \begin{equation}
L_{i}(y_{k}^{(i)})=0.\label{L-zero}\end{equation}
 The inverse statement is not true, in the sense that there can be
points where $L_{i}$ vanishes without corresponding to some zero.

We do not deal here with the numerical analysis of the TBA equations
found above. Details can be found in \cite{FPR} and will be an important
part of our paper in preparation \cite{FPRII} where we explore the
other integrable flows of the TIM. Generalizations to higher minimal
models and to other rational CFT's should follow the same lines, although
they can become more technically involved.

\section*{Acknowledgments}

FR thanks the organizers of the Landau meeting for giving him the
opportunity to present these results. This work was supported in part
by the European Network EUCLID, contract no. HPRN-CT-2002-00325, by
INFN Grant TO12 and by the Australian Research Council.


\begin{thebibliography}{10}
\bibitem{TIM}K. Graham, I. Runkel and G.M.T. Watts, hep-th/0010082;
P. Dorey, M. Pillin, A. Pocklington, I. Runkel, R. Tateo, G.M.T.
Watts, hep-th/0010278. Talks given
at \textit{Nonperturbative Quantum Effects, Budapest 2000}. 
\bibitem{TCSA}V.P. Yurov and Al.B. Zamolodchikov, Int. J. Mod. Phys. \textbf{A5}
(1990) 3221. 
\bibitem{DPTW}P. Dorey, A. Pocklington, R. Tateo and G.M.T. Watts, Nucl. Phys. \textbf{B525}
(1998) 
\bibitem{gthm}I. Affleck and A. Ludwig, Phys. Rev. Lett. \textbf{67} (1991) 161.
641
\bibitem{YangYang}C.N. Yang and C.P. Yang, J. Math. Phys. \textbf{10} (1969) 1115.
\bibitem{Zam90}Al.B. Zamolodchikov, Nucl. Phys. \textbf{B342} (1990) 695. 
\bibitem{Cardy}J.L. Cardy, Nucl. Phys. \textbf{B324} (1989) 481. 
\bibitem{chim}L. Chim, J. Math. Phys. A\textbf{11} (1996) 4491.
\bibitem{Affleck}I. Affleck, J. Phys. A\textbf{33} (2000) 6473.
\bibitem{HardSq}R.J. Baxter, J. Phys. \textbf{A13} (1980) L61; R.J. Baxter, {}``Exactly
Solved Models in Statistical Mechanics'', Academic Press, London,
1982; R.J. Baxter and P.A. Pearce, J. Phys. \textbf{A15} (1982) 897;
J. Phys. \textbf{A16} (1983) 2239. 
\bibitem{ABF}G.E. Andrews, R.J. Baxter and P.J. Forrester, J. Stat. Phys. \textbf{35}
(1984) 193. 
\bibitem{BP}R.E. Behrend and P.A. Pearce, J. Phys. A \textbf{29} (1996), 7827;
Int. J. Mod. Phys. \textbf{B11} (1997) 2833; J. Stat. Phys. \textbf{102}
(2001) 577. 
\bibitem{OPW}D.L. O'Brien, P.A. Pearce and S.O. Warnaar, Nucl. Phys. \textbf{B501},
773 (1997). 
\bibitem{BPO}R.E. Behrend, P.A. Pearce and D.L. O'Brien, J. Stat. Phys. \textbf{84},
1 (1996).
\bibitem{bauer-saleur}M. Bauer and H. Saleur, Nucl. Phys. \textbf{B320} (1989) 591
\bibitem{FinChar}E. Melzer, Int. J. Mod. Phys. \textbf{A9} (1994) 1115; A. Berkovich,
Nucl. Phys. \textbf{B431} (1994) 315. 
\bibitem{FPR}G. Feverati, P.A. Pearce and F. Ravanini, Phys. Lett. \textbf{B534}
(2002) 216
\bibitem{FPRII}G. Feverati, P.A. Pearce and F. Ravanini, in preparation\end{thebibliography}
\end{document}